\documentclass[12pt]{article}
\usepackage{aas_macros}
\usepackage{times}
\usepackage{geometry}
\usepackage[utf8]{inputenc}
\usepackage{enumitem,amssymb}
\usepackage{ragged2e}
\usepackage{graphicx}
\usepackage{fancyhdr}
\usepackage{hyperref}
\usepackage[
    natbib=true,
    style=numeric,
    sorting=none
]{biblatex}
\usepackage{color, colortbl}
\newlist{thematic}{itemize}{8}
\setlist[thematic]{label=$\square$}
\usepackage{pifont}

\newcommand\farcs{\mbox{$.\!\!^{\prime\prime}$}}%

\begin{document}

%% Page 1: Title, technical details, observing description, 1 page limit for this section

\raggedright
\huge{
A Striking First Impression: CGI Commissioning Observations of the \\AB Aurigae Protoplanetary System}
\linebreak
\large

Thayne Currie$^{1,2}$,
Kellen Lawson$^{3}$, 
Erica Dykes$^{1}$, 
Mona El Morsy $^{1}$
\\
1 Department of Physics and Astronomy, University of Texas-San Antonio, San Antonio, TX, USA \\
2 Subaru Telescope, National Astronomical Observatory of Japan, 
Hilo, HI, USA\\
3 NASA Goddard Spaceflight Center, Greenbelt, MD, USA

\justify{
\textbf{Abstract:}  
For one of the first set of Roman Coronagraph project images, we propose to target AB Aurigae.   AB Aurigae is a complex and visually stunning system, surrounded by a gas rich protoplanetary disk showing numerous spiral arms, an enigmatic embedded protoplanet (AB Aurigae b) at 0\farcs{}6 separation, and hints of potential additional sites of planet formation.   Even a marginally-successful dark hole generation (e.g. 10$^{-5}$--10$^{-6}$ contrast) with CGI would yield a vastly improved view of AB Aur b at optical wavelengths where current ground-based and HST data struggle to yield a high SNR detection and parameters (astrometry, photometry) unbiased by processing artifacts.  Total intensity imaging and polarimetry together will provide new constraints on the disk's dust properties and the range of emission sources for AB Aur b. 
AB Aur images with the Roman Coronagraph will provide a striking, inspiring demonstrations of the instrument's power and promise for detecting fainter planets and disks.}

\pagebreak
%replace \square with \boxtimes to select type of observation.

\noindent \textbf{Type of observation:} \\$\square$   Technology Demonstration\\
$\boxtimes$ Scientific Exploration\\\\

\noindent \textbf{Scientific / Technical Keywords:}  
companion (substellar), companion (exoplanet), self-luminous, high contrast performance
%\textcolor{blue}{Please choose one or more keywords from the following list:
%disk, companion (exoplanet), companion (substellar), self-luminous, reflected light, other science, observing strategy, post-processing, control algorithm, high contrast performance} \linebreak
\\

\noindent \textbf{Required Detection Limit:}  
%\textcolor{blue}{Please estimate the detection limit necessary for your observation.
%Note: the higher your contrast requirement, the less likely the observation will be feasible in the baseline observing period.  Observations that require detection limits better than 10$^{-8}$ may not be possible.  Observations requiring detection limits from 10$^{-5}$ to 10$^{-6}$ may be appropriate for ``first-look” Coronagraph observations early in the mission, prior to the Coronagraph achieving its full high-contrast performance. }
%\linebreak
%\begin{center}
\begin{tabular}{| c | c | c | c | c |}
\hline
$\geq$10$^{-5}$ & 10$^{-6}$ & 10$^{-7}$ & 10$^{-8}$ & 10$^{-9}$ \\ \hline
 x& x&  & & \\ \hline
\end{tabular}
%\end{center}

% \noindent \textbf{Estimate of Observation Length (hours):}  
% %Please estimate the number of hours of observations (including overheads) to achieve your scientific and technical goals).
% \linebreak

% \noindent \textbf{Estimate of Risk Level:}  
% %Please state whether the observation is low, medium, or high risk.
% \linebreak

%Please choose from one of the following supporting or best effort modes (delete or comment out rows from table that you will not use).  %NB: you can select non-supported modes as well, but these modes will likely not be attempted during "first-look" / commissioning and may not be available.

\vspace{0.5cm}
\textbf{Roman Coronagraph Observing Mode(s):} 
%\textcolor{blue}{Please indicate your desired observing mode(s). You may choose from one of the following supported or best effort modes (delete or comment out rows from table that you will not use). You may also choose non-coronagraphic imaging in any of the four filters. You can edit the table to request non-supported modes as well, but these modes will likely not be attempted during the initial ``first-look"/commissioning period and may not be available during the baseline observation phase.  The CPP still solicits observing ideas in non-supported modes as part of a potential further extended observing period with the Roman Coronagraph.}

%\linebreak 
\begin{tabular}{| c | c | c | c | c |}
\hline
Band &   Mode & Mask Type & Coverage & Support \\ \hline \hline

%% Band 1, narrow field imaging and polarimetry
1, 575 nm &   Narrow FoV & Hybrid Lyot & 360$^{\circ}$ & Required (Imaging), \\
 & Imaging &  &  & Best Effort (Polarimetry) \\ \hline 

%% Band 1, wide field imaging and polarimetry
1, 575 nm &  Wide FoV & Shaped Pupil & 360$^{\circ}$ & Best Effort  \\ 
 &  Imaging &  &  & (Imaging and polarimetry) \\
\hline

%% Band 2, spectroscopy
%2, 659 nm &  Slit + R$\sim$50  & Shaped Pupil & 2x65$^{\circ}$ & Best Effort \\ 
% &    Prism Spectroscopy &  &  & \\  \hline

%% Band 3, spectroscopy
%3, 730 nm &  Slit + R$\sim$50  & Shaped Pupil & 2x65$^{\circ}$ & Best Effort \\ 
% &    Prism Spectroscopy &  &  & \\  \hline

%% Band 4, wide field imaging and polarimetry
4, 825 nm & Wide FoV  & Shaped Pupil & 360$^{\circ}$ & Best Effort  \\ 
 & Imaging &  & &  (Imaging and polarimetry)  \\ \hline
\end{tabular}

\justify{
\begin{center}
\begin{tabular}{| c | c | c | c | c |}
\hline
Name &  host star & detection & separation (") & description \\
  & V mag. & limit & (or extent)  & \\ \hline \hline
  AB Aurigae & 7.05 & 10$^{-4}$--10$^{-6}$ & 0.15--1.4 & disk and protoplanet\\
  %PSF reference star & $<$3 & -- & & PSF reference \\
  %51 Eri b & 5.4 & 7$\times$10$^{-8}$ & 0.23 & self-luminous exoplanet\\
  %HR 8799 e & 5.4 & 7$\times$10$^{-8}$ & 0.23 & self-luminous exoplanet\\
 % AF Lep b & 6.3 & 1$\times$10$^{-8}$ & 0.23 & self-luminous exoplanet\\
  %HD 63754 B & 6.5 & 7$\times$10$^{-8}$ & 0.48 & self-luminous exoplanet\\ % to 2$\times$10$^{-7}$
  %HIP 54515 b & 6.8 & 3$\times$10$^{-6}$ & 0.23 & self-luminous exoplanet\\
  %HD 206893 B & 6.7 & 5$\times$10$^{-9}$ & 0.20 & self-luminous brown dwarf/exoplanet\\
  \hline
\end{tabular}
\end{center}

}

% Optional Technical Questions regarding example targets.
\justify{
\noindent \textbf{Optional Questions:}
%\textcolor{blue}{Please complete only if you have specific targets in mind -- fine to skip this if you are proposing for a generic category of targets.}
}

\justify{
\noindent \textbf{Are any example targets binary systems in the Washington Double Star survey or other binary survey?} 
No

}

\justify{
\noindent \textbf{Do any of your example Hybrid Lyot coronagraphic target stars have angular diameters $>$ 2 mas?}
No
}

\pagebreak

%%% Page 3: Description of Scientific / Technological Goals, 1-page limit for this section

\justify{
% 1 page of Scientific / Technical Justification, including an estimate of time needed.

\begin{figure}[!h]
\centering
 \includegraphics[width=0.9\textwidth,clip]{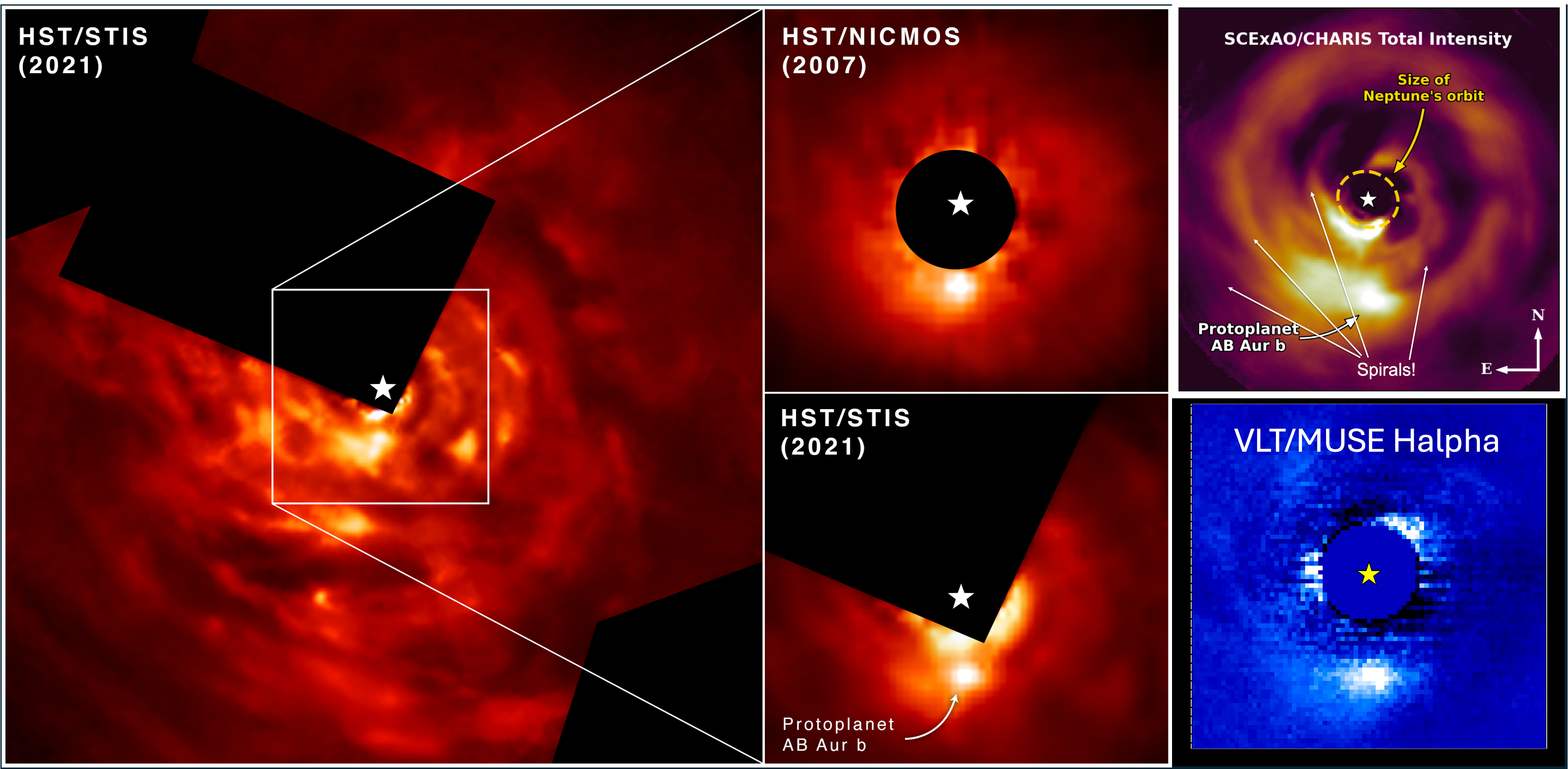}
    \vspace{-0.15in}
    \caption{Gallery of AB Aurigae images.  (left) Detection of the AB Aur disk with HST/STIS, where insets show the detection of AB Aur b with NICMOS and STIS.  (top-right) Detection of the AB Aur b protoplanet and numerous spiral arms with SCExAO/CHARIS in the near-infrared.  (bottom-right) Detection of AB Aur b in $H_{\rm \alpha}$ medium-resolution high-contrast spectroscopy with VLT/MUSE.
    }
\end{figure}

\textbf{\Large Anticipated Technology / Science Objectives:}

Direct images of \textit{protoplanets} within their natal protoplanetary disks provide key insights into gas giant planet formation \citep{Currie2023b}, including the origin of many of the $\sim$25 superjovian directly-imaged exoplanets \citep[e.g.][]{Marois2008,Chauvin2017,Currie2023a,Currie2025}.  A subset of protoplanetary disks around stars more than twice as massive as the Sun show complex spiral density waves that may be signatures of dynamical perturbations from massive, moderate-to-wide separation planets \citep{Benisty2023}.  Of these systems, the A0-type Taurus member AB Aurigae is arguably one of the best studied, hotly debated, and most elusive.

AB Aurigae is a complex and visually stunning system: in its highest-quality scattered-light images, its disk superficially resembles a spiral galaxy more than a typical protoplanetary disk \citep{Boccaletti2020,Currie2022,Lawson2022,Dykes2024,Speedie2024} (Fig. 1).      Moreover, SCExAO/CHARIS and HST/NICMOS and STIS data identify AB Aur b, an embedded protoplanet orbiting counterclockwise from the star ($\rho$ = 0\farcs{}6, $\sim$100 au), at a location consistent with the predicted position of a protoplanet driving CO gas spirals and located and within the submm-imaged dust cavity \citep{Tang2012,Tang2017}. 
Most recently, VLT/MUSE detects AB Aur b in $H_{\rm \alpha}$, only the second such system with an $H_{\rm \alpha}$ detection \citep{Currie2025a}, and the first showing an inverse P Cygni profile consistent with infalling material.

Currently, the main weakness in understanding AB Aurigae is in the optical, where high-quality AO corrections are very challenging.  While HST/STIS delivers excellent high contrast, its PSF is not well sampled.   The scientific return of the only other space-based option (WFC3) is plagued by a lack of a coronagraph, relatively poor contrasts, and complex self-subtraction biasing of post-processed data which are incredibly hard to calibrate \citep{Currie2025a}.

 Even a marginally-successful dark hole generation (e.g. 10$^{-5}$--10$^{-6}$ contrast) with CGI would yield a vastly improved view of AB Aur b at optical wavelengths where current ground-based and HST data struggle to yield a high SNR detection and parameters (astrometry, photometry) unbiased by processing artifacts.   A combination of narrow field and wide-field Band 1 imaging with the hybrid Lyot coronagraph and wide-field Band 4 imaging with the shaped pupil is sufficient to cover a full 360$^{o}$ view of AB Aur out to 1\farcs{}4.  Total intensity imaging and polarimetry together will provide new constraints on the disk's dust properties and the range of emission sources for AB Aur b \citep{Dykes2024}. 

\textbf{Is this observation appropriate for ``first-look" / commissioning ($<$3 months after launch), the observation phase ($<$ 18 months after ``first-look" / commissioning), or a potential extended observing phase?}: 
This is an ideal ``first-look" observation for the Roman Coronagraph.  The disk and protoplanet are so bright that they are visible in raw images with SCExAO/CHARIS and HST/STIS.  Deep contrasts ($<$10$^{-6}$) are not necessary but could newly reveal planet-related features.  
%This program also provides a useful first examination of CGI's performance for V = 7 stars before its performance is better tuned to achieve TTR5 and image companions around many V = 5--7 stars.

%\pagebreak

%% Page 4: Observation Description, 1-page limit for this section, 1 page limit for this section
\textbf{\Large Observing Description:}\\
The program goal requires Band 1/575 nm imaging and polarimetry with the hybrid Lyot coronagraph -- in both narrow and wide fields -- and Band 4/825 nm wide-field imaging and polarimetry with the shaped pupil. The observations should follow the ``standard typical observing sequence" described on page 36 of the most recent CGI white paper slide deck that enables both angular and reference star differential imaging (ADI, RDI)\footnote{Coronagraph$\_$CPP$\_$WPoverview2025$\_$8July2025}.  Each "visit" consists of dark hole digging on a bright nearby PSF reference, PSF reference observations, two sets of +/- telescope rolls ($\Delta$$\theta$= +/- 15$^{o}$) on the target star, followed by a second set of PSF reference observation.  
The combined HLC and SPC data will cover 0\farcs{}15--0\farcs{}45 and 0\farcs{}4--1\farcs{}4: i.e. the full field of view over which AB Aur displays some of its most spectacular features.

There are a couple of reasonable PSF reference stars near AB Aur that could work: e.g. eps Ori and $\gamma$ Ori.  As shown in Fig. 2, there are windows of time where the pitch angle difference from the reference star is less than 3--5 $^{o}$.

%The rollfollow exactly what is baselined to fulfill TTR5
%\textcolor{blue}{Please include a description (up to 1 page) of the planned observing idea, with an estimate of the observing time necessary to achieve your goals.  This is the portion of the white paper where you can include details on: the choice of the coronagraph mask and bandpass to be used, required exposure sequence, required detection limit and detection S/N, number and spacing of visits, whether angular differential imaging (ADI) or reference differential imaging (RDI) or both are needed, as appropriate for your particular observing idea.}

\textbf{Estimate of Time Needed}:  The time required to complete this program is fairly minimal.   AB Aur b is easily detectable with HST and SCExAO/CHARIS already (Fig. 2).   For a goal of a 10-$\sigma$ contrast of 10$^{-6}$ (set by the speckle floor, not the bright disk), we obtain exposure times of 0.4--0.5 hours in Band 1 and 0.6--0.7 hours in Band 4.  Assuming the nominal overhead of $\sim$ 1.7 (14 hours integration time in 24 hours clock time), both bands in both total and polarized intensity can be done in just a few hours.
\vspace{0.2cm}

%\pagebreak

%% Page 5: Up to 3 figures, 1 page limit for this section

\begin{figure}[!h]
\centering
 \includegraphics[width=0.3\textwidth,clip]{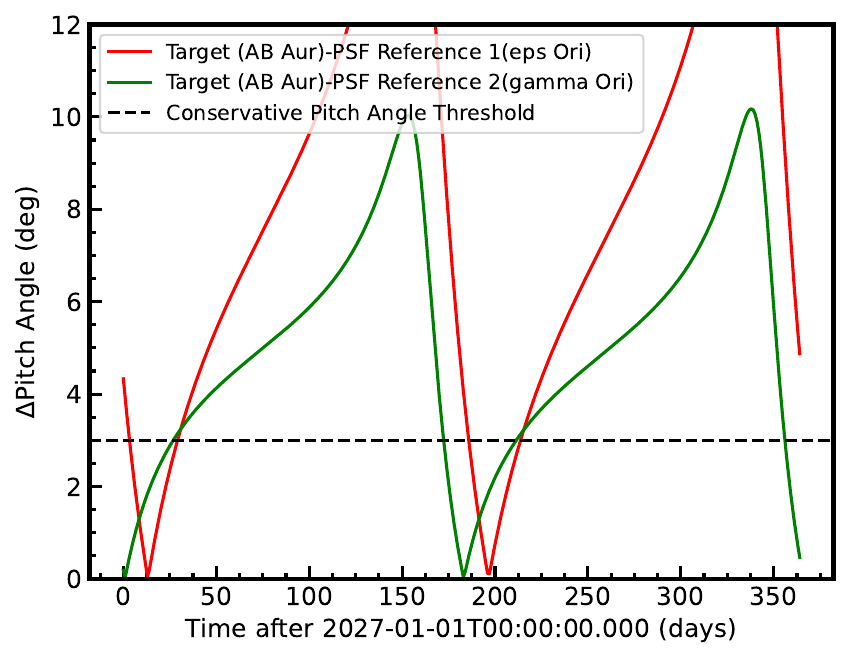}
      \includegraphics[width=0.31\textwidth,clip]{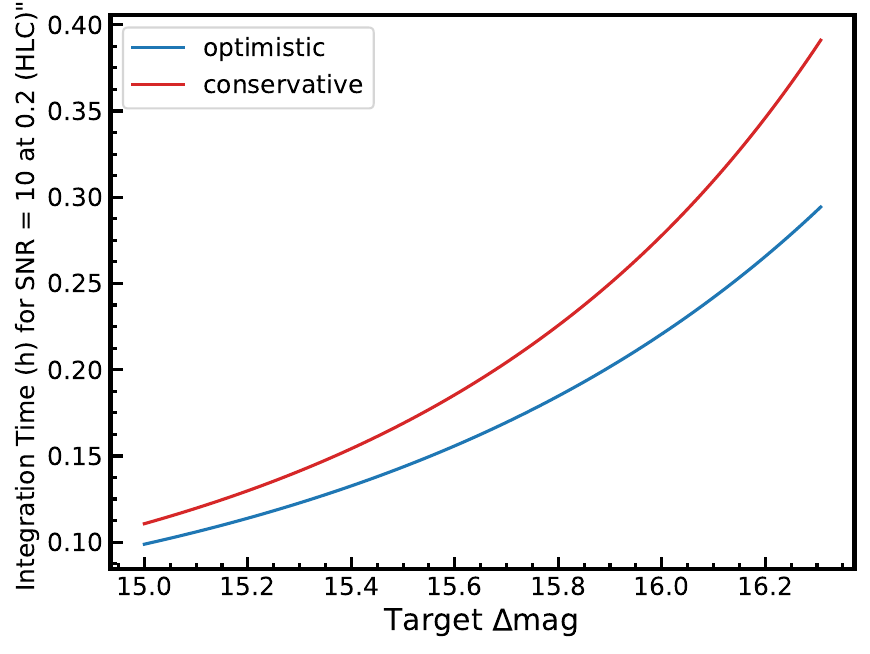}
      \includegraphics[width=0.3\textwidth,clip]{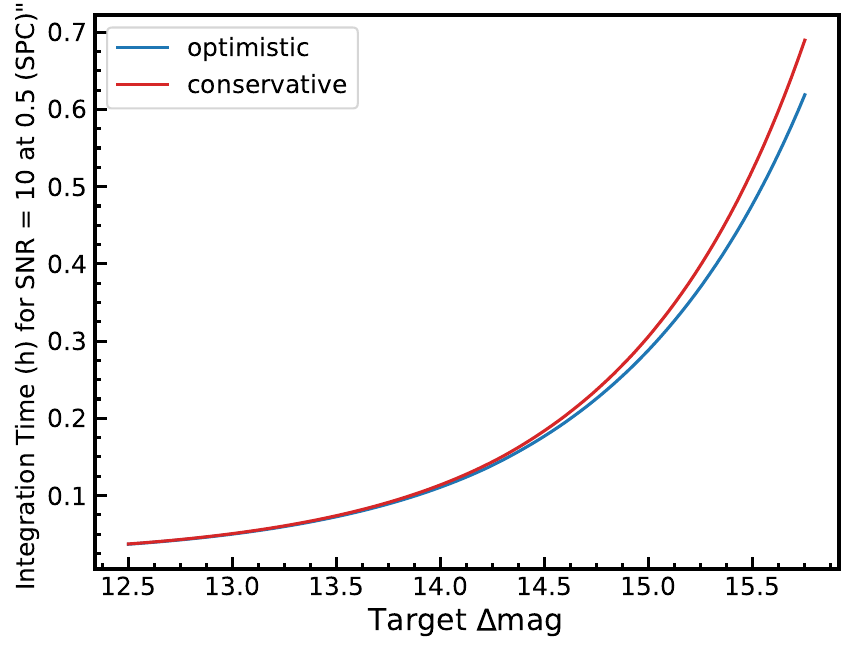}
    \vspace{-0.15in}
    \caption{(left) Relative Pitch Angle between AB Aur and two potential reference stars during the calendar year of 2027.  The nominal "good" performance cutoff at 3$^{o}$ is shown but again this program does not require TTR5-level contrast performance.   (middle, right) Time to reach SNR = 10 for a given contrast.  Note that AB Aur b roughly has a contrast of 10$^{-4}$ at optical wavelengths.
    %.We assume a 1, 10x, 100x, and 10$^{5}$ solar system exo-zodi level for HIP 54515, HIP 99770, HR 8799, and $\beta$ Pic b, respectively.  
    }
\end{figure}

\pagebreak

\printbibliography

\end{document}